\documentclass[10pt, conference]{IEEEtran}
\IEEEoverridecommandlockouts

\usepackage{cite}
\usepackage{amsmath,amssymb,amsfonts}

\usepackage{amsthm}

\usepackage[linesnumbered, ruled]{algorithm2e}
\SetNlSty{}{}{:}   
\let\oldnl\nl
\newcommand{\nlnonumber}{\renewcommand{\nl}{\let\nl\oldnl}}

\usepackage{url}

\usepackage{graphicx}
\usepackage{textcomp}
\usepackage{xcolor}
\usepackage{xspace}
\usepackage{tabularx}
\usepackage{wrapfig}
\usepackage{pbox}
\usepackage{bbding}
\usepackage{textcomp}
\usepackage{xcolor}

\usepackage[linesnumbered, ruled]{algorithm2e}

\usepackage{bbm}
\usepackage{dblfloatfix}
\usepackage{pifont}
\usepackage{bm}
\usepackage{enumitem}
\usepackage{svg}
\usepackage{comment}

\usepackage{booktabs}
\usepackage{diagbox}

\usepackage{rotating}
\usepackage{tabularray}

\def\BibTeX{{\rm B\kern-.05em{\sc i\kern-.025em b}\kern-.08em
    T\kern-.1667em\lower.7ex\hbox{E}\kern-.125emX}}

\begin{document}

\title{Toward Efficient Sensing in Multi-Device ISCC by Removing Frequency Domain Redundancy}

\author{
\IEEEauthorblockN{Ziqi Ye$^{1}$, Yinghui He$^{1\dagger}$,  Weiwei Chen$^{1}$, Guanding Yu$^{1}$, \IEEEmembership{Senior Member,~IEEE}, and Rui Zhang$^{2}$, \IEEEmembership{Fellow,~IEEE}}
\IEEEauthorblockA{$^{1}$College of Information Science and Electronic Engineering, Zhejiang University, Hangzhou, China}
\IEEEauthorblockA{$^{2}$Department of Electrical and Computer Engineering, National University of Singapore, Singapore}
\IEEEauthorblockA{Email: \{yeziqi, 2014hyh, 22331145, yuguanding\}@zju.edu.cn, elezhang@nus.edu.sg, $\dagger$~Corresponding author}
}

\maketitle
\begin{abstract}
Integrated sensing, communication, and computation (ISCC) is envisioned as a key enabler for intelligent services in future wireless networks. However, in multi-device ISCC systems, directly offloading full orthogonal frequency division multiplexing (OFDM) sensing data to the edge may incur excessive overhead, thereby limiting sensing performance under practical resource constraints. In this paper, we propose a subcarrier selection-based sensing framework for multi-device ISCC systems, where frequency-domain redundancy in OFDM sensing data is removed during local preprocessing to reduce sensing data transmission and processing overhead. Based on the proposed framework, we establish analytical models for sensing accuracy, delay, and energy consumption, and formulate a sensing accuracy maximization problem under practical resource constraints. To solve this problem, we develop an alternating direction method of multipliers (ADMM)-based algorithm. Experiments on commodity wireless devices validate the effectiveness of the proposed framework and show that it consistently outperforms three baseline schemes under various resource constraints.
\end{abstract}

\section{Introduction}
As 6G networks evolve toward greater intelligence and ubiquitous connectivity, wireless systems are expected to support not only reliable communications but also human activity sensing~\cite{liu2022integrated}. In this context, integrated sensing and communication (ISAC) enables applications such as human detection, localization, and activity recognition by jointly supporting sensing and communication~\cite{wei2023integrated, ye2026practical, zhang2025intelligent, he2025task}. Many emerging applications further require timely processing of sensed information for intelligent decision making, which drives the evolution from ISAC to integrated sensing, communication, and computation (ISCC)~\cite{wen2023taskair, yang2020federated}. As a promising architecture for edge intelligence, ISCC further integrates sensing, wireless transmission, and computation in a unified framework~\cite{wen2024survey}. In multi-device scenarios, however, each Internet of Things (IoT) device continuously collects a large volume of orthogonal frequency division multiplexing (OFDM) sensing data over multiple frequency subcarriers and offloads them to the edge for inference. Directly transmitting full sensing data from all devices may incur excessive communication and computation overhead, thereby limiting sensing performance under practical latency, device energy, and edge computing resource constraints~\cite{li2022joint}. Therefore, efficient local preprocessing is needed to reduce unnecessary sensing overhead while preserving critical sensing information.

Existing studies have shown that local preprocessing before edge offloading is an effective way to improve sensing efficiency. For example, schemes based on action detection can avoid unnecessary transmission for static windows~\cite{he2024integrated}, while recent studies have further exploited temporal characteristics to reduce the upload of less informative sensing windows~\cite{chen2025sensing}. These studies demonstrate that reducing redundant sensing data before transmission is important for improving the overall efficiency of ISCC systems. However, existing efforts mainly focus on how to reduce temporal redundancy. In OFDM-based wireless sensing, adjacent subcarriers are close in frequency and thus often carry highly correlated activity-related information, which implies that substantial redundancy may also exist in the frequency domain. Once a sensing window is selected for transmission, directly offloading OFDM sensing data from all subcarriers may incur considerable overhead without providing commensurate sensing gains. Therefore, how to remove redundancy across subcarriers remains an important yet underexplored problem in multi-device ISCC systems.

Motivated by the above, in this paper, we propose a subcarrier selection-based sensing framework for multi-device ISCC systems. In the proposed framework, each device first performs lightweight local preprocessing and then retains only representative subcarriers before offloading sensing data to the edge, thereby effectively reducing frequency-domain redundancy in multi-subcarrier OFDM sensing data. Based on this framework, we establish analytical models for sensing accuracy, delay, and energy consumption, and formulate a sensing accuracy maximization problem under practical constraints on delay, energy, and edge computing. To solve the optimization problem, we further develop an alternating direction method of multipliers (ADMM)-based algorithm to jointly optimize local sensing parameters and global resource allocation. Experiment results on commodity wireless devices validate the effectiveness of the proposed framework. In particular, they show that the proposed subcarrier selection mechanism can substantially reduce sensing overhead with limited accuracy loss, while the overall scheme consistently outperforms three baseline schemes under different constraints.

\begin{figure*}[t]
    \centering
    \begin{minipage}[t]{0.4\textwidth}
        \centering
        \includegraphics[height=3.6cm,keepaspectratio]{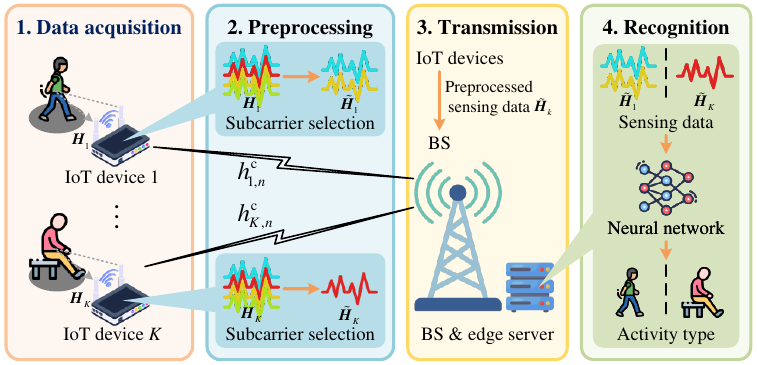}
        \vspace{-4ex}
        \caption{The considered ISCC system and the proposed sensing framework.}
        \label{fig:system}
    \end{minipage}\hfill
    \addtocounter{figure}{1}
    \begin{minipage}[t]{0.25\textwidth}
        \centering
        \includegraphics[height=3.4cm,keepaspectratio]{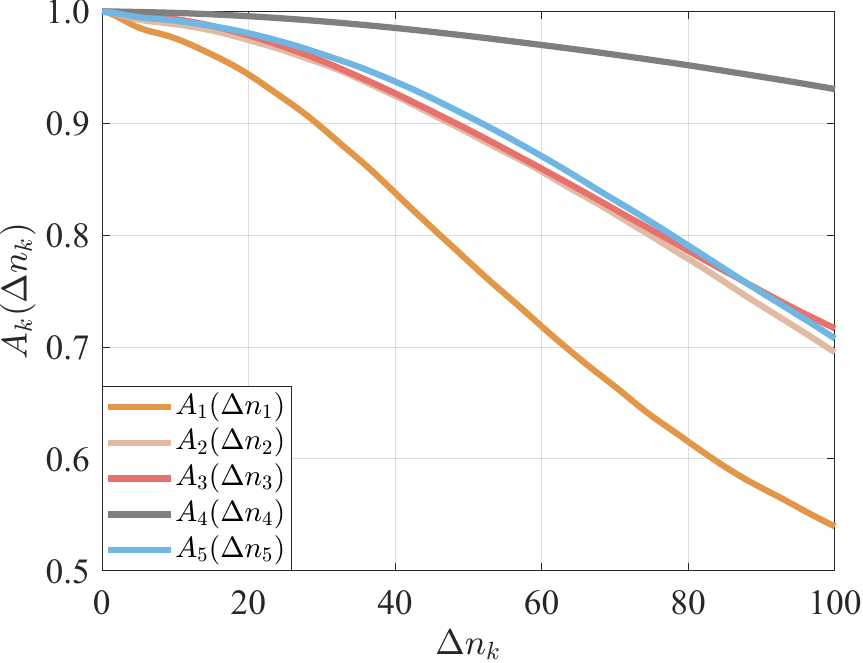}
        \vspace{-4ex}
        \caption{$A_k(\Delta n_k)$ vs. $\Delta n_k$ for five IoT devices.}
        \label{fig:Autocorrelation}
    \end{minipage}\hfill
    \begin{minipage}[t]{0.25\textwidth}
        \centering
        \includegraphics[height=3.6cm,keepaspectratio]{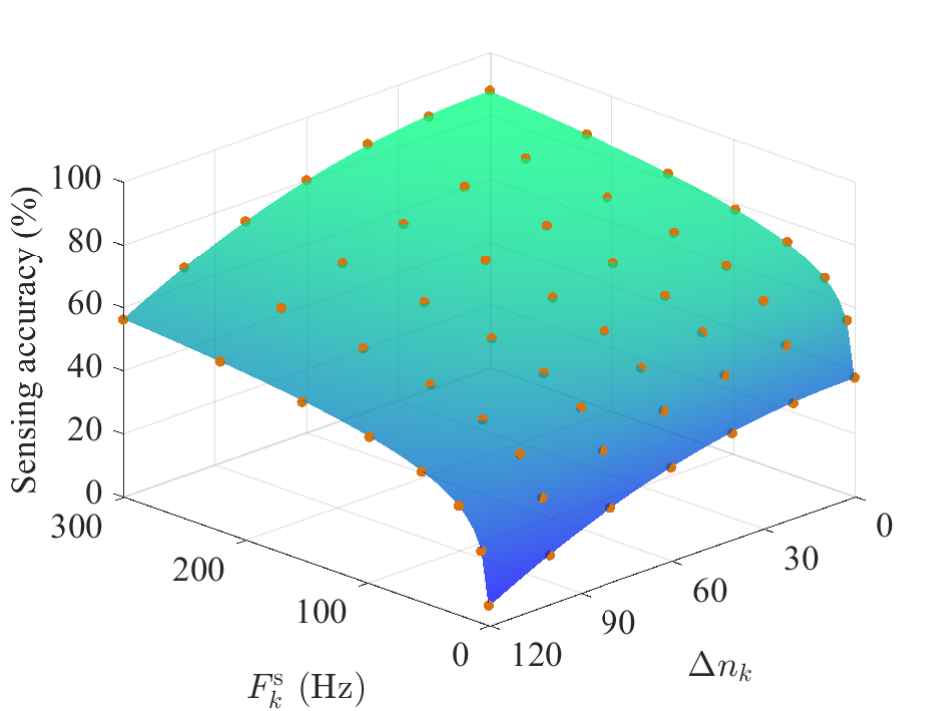}
        \vspace{-4ex}
        \caption{Sensing accuracy vs. sampling rate and subcarrier selection interval.}
        \label{fig:acc_F_Delta_n}
    \end{minipage}
    \vspace{-2.5ex}
\end{figure*}

\setcounter{figure}{1}
\begin{figure}[t]
    \centering
    \includegraphics[width=0.8\linewidth]{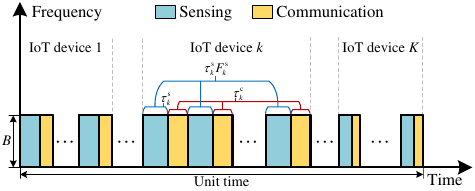}
    \vspace{-2ex}
    \caption{Time-frequency resource allocation for the considered ISCC system.}
    \vspace{-3ex}
    \label{fig:time_frequency_resource}
\end{figure}
\setcounter{figure}{4}

The rest of this paper is organized as follows. Section~\ref{sec:sys} introduces the system model and the proposed sensing framework. Section~\ref{sec:per_ana} presents the performance analysis and formulates the sensing accuracy maximization problem. Section~\ref{sec:algorithm} develops the ADMM-based algorithm. Section~\ref{sec:evalation} provides the evaluation results. Finally, Section~\ref{sec:con} concludes this paper.

\section{System Model and Sensing Framework}
\label{sec:sys}
\subsection{ISCC System}
\label{sec:iscc_sys}

We consider a multi-device ISCC system consisting of one base station (BS) equipped with an edge server and $K$ IoT devices, where each device is equipped with a dual-functional OFDM transceiver. As shown in Fig.~\ref{fig:system}, each IoT device continuously senses its local environment, performs lightweight local preprocessing, and then offloads the resulting OFDM sensing data to the BS for edge-side activity recognition. In this work, we propose a device-side subcarrier selection module to remove redundancy across subcarriers before data offloading, thereby reducing transmission and processing overhead. This module constitutes the key new design in the proposed sensing framework.

To support multi-device sensing and communication, the BS adopts time division multiple access (TDMA) for device access. As illustrated in Fig.~\ref{fig:time_frequency_resource}, during its allocated time, each IoT device alternates between sensing and uplink transmission while occupying the full system bandwidth $B$. This design enables seamless integration of sensing, communication, and edge intelligence under practical resource constraints.

\subsection{Sensing Procedure with Subcarrier Selection}
\label{sec:device_edge_procedure}

For the $n$-th subcarrier of device $k$, let $x_{k,n}$ and $y_{k,n}$ denote the transmitted signal and the corresponding echo, respectively. The OFDM sensing signal is then defined as $h_{k,n}={y_{k,n}}/{x_{k,n}}$. Over a sensing window of duration $T^{\mathrm{s}}$ and sensing sampling rate $F_k^{\mathrm{s}}$, device $k$ collects $T^{\mathrm{s}}F_k^{\mathrm{s}}$ samples on each subcarrier and forms the sensing data matrix $\bm{H}_k=[\bm{h}_{k,1},\bm{h}_{k,2},\ldots,\bm{h}_{k,N}]^T
\in \mathbb{C}^{N\times (T^{\mathrm{s}}F_k^{\mathrm{s}})}$,
where $\bm{h}_{k,n}$ denotes the data sequence of the $n$-th subcarrier. Let $\tau_k^{\mathrm{s}}$ denote the time required for one sensing cycle. Then, the sensing operation of device $k$ occupies a time fraction $\tau_k^{\mathrm{s}}F_k^{\mathrm{s}}$ within unit time.

Directly offloading $\bm{H}_k$ to the BS would incur substantial transmission and computation overhead. To address this issue, each device performs local preprocessing before data transmission, including the action detection module~\cite{he2024integrated} and, more importantly, the proposed subcarrier selection module. The latter is designed to remove redundant sensing information across subcarriers before offloading.

At the beginning of local preprocessing, the discrete Fourier transform (DFT) of $\bm{h}_{k,n}$ is computed as
$w_{k,n}[f]
=
({1}/{\sqrt{T^{\mathrm{s}}F_k^{\mathrm{s}}}})
\sum_{m=1}^{T^{\mathrm{s}}F_k^{\mathrm{s}}}
h_{k,n}[m]
\exp\!\left(
-{j2\pi fm}/({T^{\mathrm{s}}F_k^{\mathrm{s}}})
\right)$,
where $f=0,\ldots,T^{\mathrm{s}}F_k^{\mathrm{s}}\!-\!1$. Focusing on the activity-related high-frequency range $[f^{\mathrm{L}},f^{\mathrm{U}}]$, the corresponding power spectrum is
$e_{k,n}[f]=|w_{k,n}[f]|^2, f\in [f^{\mathrm{L}},f^{\mathrm{U}}]$.
For the action detection module, according to~\cite{he2024integrated}, device $k$ computes the average high-frequency power on the first subcarrier as
$P_k=(1/{T^{\mathrm{s}}F_k^{\mathrm{s}}})
\sum_{f=f^{\mathrm{L}}}^{f^{\mathrm{U}}} e_{k,1}[f]$.
If $P_k<\delta_k$, the target is regarded as static and no sensing data are offloaded. Otherwise, the target is regarded as dynamic, and the proposed subcarrier selection module is activated.

The subcarrier selection module extracts a reduced sensing data matrix $\tilde{\bm{H}}_k$ from $\bm{H}_k$ by removing redundant subcarriers. The key idea behind our design is that adjacent subcarriers often carry correlated sensing information, so retaining only representative subcarriers can significantly reduce overhead with limited information loss. To this end, we characterize the redundancy among subcarriers through the activity-related high-frequency power spectrum. Specifically, we define
$\bm{e}_{k,n}
=
[e_{k,n}[f^{\mathrm{L}}],\ldots,e_{k,n}[f^{\mathrm{U}}]]^H
\in \mathbb{R}^{(f^{\mathrm{U}}-f^{\mathrm{L}}+1)\times 1}$,
and the subcarrier similarity function
\begin{equation}
A_k(\Delta n_k)
=
\frac{1}{N-\Delta n_k}
\sum_{n=1}^{N-\Delta n_k}
\frac{\bm{e}_{k,n}\cdot \bm{e}_{k,n+\Delta n_k}}
{\|\bm{e}_{k,n}\|\,\|\bm{e}_{k,n+\Delta n_k}\|},
\end{equation}
which measures the average cosine similarity between subcarriers separated by $\Delta n_k$ over the entire bandwidth. Fig.~\ref{fig:Autocorrelation} shows that $A_k(\Delta n_k)$ typically decreases monotonically with $\Delta n_k$. This monotonicity indicates that, as the subcarrier spacing increases, the average similarity between subcarriers gradually decreases, and thus the redundancy among them becomes weaker. Therefore, $A_k(\Delta n_k)$ can be used as a proxy for the amount of preserved sensing information when one subcarrier is retained every $\Delta n_k+1$ subcarriers. Given a similarity threshold $\eta_k$, only subcarrier intervals satisfying $A_k(\Delta n_k)\ge \eta_k$ are admissible. In this way, $\eta_k$ excludes overly aggressive selection intervals and defines a low-information-loss region for subcarrier selection, while the final $\Delta n_k$ will be jointly optimized later together with the other system parameters. Under a given admissible interval $\Delta n_k$, one subcarrier is retained every $\Delta n_k+1$ subcarriers, and the reduced sensing data matrix is given by $\tilde{\bm{H}}_k\in \mathbb{C}^{\tilde N_k\times (T^{\mathrm{s}}F_k^{\mathrm{s}})}$, where
$\tilde N_k=\left\lceil {N}/({\Delta n_k+1})\right\rceil$.

After local preprocessing, device $k$ offloads $\tilde{\bm{H}}_k$ to the BS for edge-side recognition. The instantaneous uplink transmission rate of device $k$ is
$R_k
=
({B}/{N})
\sum_{n=1}^{N}
\log_2
(
1+{|h_{k,n}^{\mathrm{c}}|^2 P_k^{\mathrm{t}}}/({N\sigma^2})
)$,
where $P_k^{\mathrm{t}}/N$ is the transmit power allocated to each subcarrier, $h_{k,n}^{\mathrm{c}}$ is the uplink channel gain on the $n$-th subcarrier, and $\sigma^2$ is the noise power. Since TDMA is adopted, let $\tau_k^{\mathrm{c}}$ denote the communication time fraction allocated to device $k$ within unit time. Then, the average uplink data rate of device $k$ within unit time is $\tau_k^{\mathrm{c}} R_k$. After receiving all $\tilde{\bm{H}}_k$, the edge server performs convolutional neural network (CNN)-based activity recognition.

\section{Performance Analysis and Problem Formulation}
\label{sec:per_ana}
In this section, we characterize the sensing accuracy, sensing delay, and energy consumption, and then formulate the sensing accuracy maximization problem.

\subsection{Sensing Accuracy}
\label{sec:sen_acc}
We consider an $I$-class activity recognition task, where the first class ($i=1$) represents the static type and the remaining classes correspond to dynamic activity types. The sensing performance is affected by the device-side action detection and subcarrier selection modules, as well as the edge-side activity recognition module. Specifically, action detection first distinguishes static and dynamic sensing windows, while for the detected dynamic windows, the proposed subcarrier selection module and the edge-side CNN further determine the final recognition performance.

Following~\cite{he2024integrated}, the performance of action detection is characterized by the miss rate $p_{k,i}^{\mathrm{o}}$ and the false positive rate $p_{k,1}^{\mathrm{l}}$. Specifically, $p_{k,i}^{\mathrm{o}}$ denotes the probability that an instance of the $i$-th dynamic type is incorrectly recognized as static, while $p_{k,1}^{\mathrm{l}}$ denotes the probability that a static instance is incorrectly recognized as dynamic. These two probabilities are determined by $F_k^{\mathrm{s}}$ and the detection threshold $\delta_k$. To ensure reliable action detection, they are required to satisfy $p_{k,i}^{\mathrm{o}} \le p^{\max}, i=2,\ldots,I$ and $p_{k,1}^{\mathrm{l}} \le p^{\max}$, where $p^{\max}$ is a predefined reliability threshold. Under these constraints, the errors introduced by action detection are sufficiently small and are thus neglected in the sensing accuracy model. Accordingly, we use the edge-side recognition accuracy for the retained dynamic sensing data to approximate the overall sensing accuracy. Let $\alpha_k(F_k^{\mathrm{s}}, \Delta n_k)$ denote the recognition accuracy when device $k$ adopts sensing sampling rate $F_k^{\mathrm{s}}$ and subcarrier selection interval $\Delta n_k$. As shown in Fig.~\ref{fig:acc_F_Delta_n}, experiment results show that $\alpha_k(F_k^{\mathrm{s}}, \Delta n_k)$ increases with $F_k^{\mathrm{s}}$ and decreases with $\Delta n_k$. Then, the sensing accuracy metric of device $k$ is given by
$A_k^{\mathrm{s}}
=
p_{k,1}^{\mathrm{m}}
+
\sum_{i=2}^{I} p_{k,i}^{\mathrm{m}} \alpha_k(F_k^{\mathrm{s}}, \Delta n_k)$,
where $p_{k,i}^{\mathrm{m}}$ denotes the occurrence probability of the $i$-th activity type and satisfies $\sum_{i=1}^{I} p_{k,i}^{\mathrm{m}}=1$.

\subsection{Delay and Energy Consumption}
\label{sec:delay_energy}
For device $k$, the sensing delay consists of sensing, local preprocessing, transmission, and edge recognition. The sensing delay is $T_k^{\mathrm{sen}} = T^{\mathrm{s}}$. Using a standard analysis based on operation counts, the local preprocessing delay is approximated by
$T_k^{\mathrm{comp}}
=
\gamma_k^{\mathrm{c,1}} T^{\mathrm{s}} F_k^{\mathrm{s}}
\log(T^{\mathrm{s}} F_k^{\mathrm{s}})
+
(
\gamma_k^{\mathrm{c,2}} N T^{\mathrm{s}} F_k^{\mathrm{s}}
\log(T^{\mathrm{s}} F_k^{\mathrm{s}})
+
\gamma_k^{\mathrm{c,3}} N^2
)
(1-p_{k,1}^{\mathrm{m}}),$
where $\gamma_k^{\mathrm{c,1}}$, $\gamma_k^{\mathrm{c,2}}$, and $\gamma_k^{\mathrm{c,3}}$ are delay parameters.
Then, let $V^{\mathrm{L}}$ denote the number of bits per matrix element and $\tilde N_k=\lceil N/(\Delta n_k+1)\rceil$ denote the number of retained subcarriers after selection. The transmission delay is
$T_k^{\mathrm{trans}}
=
{
\tilde N_k T^{\mathrm{s}} F_k^{\mathrm{s}} V^{\mathrm{L}} (1-p_{k,1}^{\mathrm{m}})
}/({
\tau_k^{\mathrm{c}} R_k
}),$
and the edge recognition delay is
$T_k^{\mathrm{CNN}}
=
{
\tilde N_k T^{\mathrm{s}} F_k^{\mathrm{s}} C^{\mathrm{e}} (1-p_{k,1}^{\mathrm{m}})
}/{
f_k^{\mathrm{e}}
}$,
where $f_k^{\mathrm{e}}$ denotes the edge computing resource allocated to device $k$ and $C^{\mathrm{e}}$ is the computational cost per matrix element. Hence, the overall sensing delay is
$T_k^{\mathrm{over}}
=
T_k^{\mathrm{sen}} + T_k^{\mathrm{comp}} + T_k^{\mathrm{trans}} + T_k^{\mathrm{CNN}}$.

Since the BS is externally powered, we only consider the energy consumption of IoT devices. The sensing, local computation, and communication energy consumption are respectively given by $E_k^{\mathrm{sen}} = P_k^{\mathrm{sen}} T^{\mathrm{s}} F_k^{\mathrm{s}}$, $E_k^{\mathrm{comp}} = P_k^{\mathrm{comp}} T_k^{\mathrm{comp}}$, and
$E_k^{\mathrm{trans}}
=
\tau_k^{\mathrm{c}} P_k^{\mathrm{t}} T_k^{\mathrm{trans}}
=
{
P_k^{\mathrm{t}} \tilde N_k T^{\mathrm{s}} F_k^{\mathrm{s}} V^{\mathrm{L}} (1-p_{k,1}^{\mathrm{m}})
}/{
R_k
}$.
Therefore, the total energy consumption of device $k$ is $E_k^{\mathrm{over}} = E_k^{\mathrm{sen}} + E_k^{\mathrm{comp}} + E_k^{\mathrm{trans}}$.

\subsection{Problem Formulation}
Our objective is to maximize the average sensing accuracy of all IoT devices under delay, energy, and edge computing resource constraints. The resulting problem is formulated as
\begin{subequations}
\label{eq:obj_fun}
\begin{eqnarray}
\!\!&\!\!\!\!\!\!\!
\max\limits_{\left\{\!\substack{F_k^{\mathrm{s}}, \Delta n_k, \\
\delta_k, \tau_k^{\mathrm{c}}, {f_k^{\mathrm{e}}}}\!\right\}}
&\!\!\!\!\!\!\!
\frac{1}{K}\!\! \sum_{k=1}^{K} \!A_k^{\mathrm{s}} \!\!=\!\! \frac{1}{K}\!\! \sum_{k=1}^{K} (p_{k,1}^{\mathrm{m}}\!\!+\!\!\!\sum_{i=2}^{I} p_{k,i}^{\mathrm{m}} \alpha_k(F_k^{\mathrm{s}}, \Delta n_k)),
~~~~~~\\
&\!\!\!\!\!\!\!\text{s.t.} &\!\!\!\!\!\!\!
p_{k,i}^{\mathrm{o}} \le p^{\max},\ i=2,\dots,I,
\label{eq:c1} \\
&\!\!\!\!\!\!\!&\!\!\!\!\!\!\!
p_{k,1}^{\mathrm{l}} \le p^{\max},
\label{eq:c2} \\
&\!\!\!\!\!\!\!&\!\!\!\!\!\!\!
A_k(\Delta n_k) \ge \eta_k,
\label{eq:c3} \\
&\!\!\!\!\!\!\!&\!\!\!\!\!\!\!
T_k^{\mathrm{over}} \!\!= \! T_k^{\mathrm{sen}} \!+ \!T_k^{\mathrm{comp}} \!+ \!T_k^{\mathrm{trans}} \!+ \!T_k^{\mathrm{CNN}} \!\le\! T_k^{\max},
\label{eq:c4} \\
&\!\!\!\!\!\!\!&\!\!\!\!\!\!\!
E_k^{\mathrm{over}} \! = \! E_k^{\mathrm{sen}} + E_k^{\mathrm{comp}} + E_k^{\mathrm{trans}} \le E_k^{\max},
\label{eq:c5} \\
&\!\!\!\!\!\!\!&\!\!\!\!\!\!\!
\sum_{k=1}^{K} (\tau_k^{\mathrm{s}} F_k^{\mathrm{s}} + \tau_k^{\mathrm{c}}) \le 1,
\label{eq:c6} \\
&\!\!\!\!\!\!\!&\!\!\!\!\!\!\!
\sum_{k=1}^{K} f_k^{\mathrm{e}} \le f^{\mathrm{e}},
\label{eq:c7} \\
&\!\!\!\!\!\!\!&\!\!\!\!\!\!\!
F_k^{\mathrm{s}} \in \mathbb{Z}^{+},
\label{eq:c8} \\
&\!\!\!\!\!\!\!&\!\!\!\!\!\!\!
\Delta n_k \in \mathbb{Z},~ 0 \le \Delta n_k \le N-1,
\label{eq:c9} \\
&\!\!\!\!\!\!\!&\!\!\!\!\!\!\!
\delta_k, \tau_k^{\mathrm{c}}, f_k^{\mathrm{e}} \ge 0,
\label{eq:c10}
\end{eqnarray}
\end{subequations}
where $T_k^{\max}$ denotes the maximum allowable sensing delay of device $k$, $E_k^{\max}$ denotes the energy budget of device $k$, and $f^{\mathrm{e}}$ denotes the total available edge computing resource. Problem~\eqref{eq:obj_fun} is difficult to solve directly due to the coupling among sensing, communication, and computing variables. In the next section, we decompose it into tractable subproblems and develop an efficient algorithm.

\section{Sensing Accuracy Maximization}
\label{sec:algorithm}

In this section, we decompose problem~\eqref{eq:obj_fun} into two subproblems and then develop an ADMM-based algorithm.

\subsection{Problem Decomposition}

To solve problem~\eqref{eq:obj_fun}, we adopt the ADMM method~\cite{wang2019global}. We first introduce a global auxiliary variable $\tau_k^{\mathrm{sum}}$ and a local auxiliary variable $\hat{f}_k^{\mathrm{e}}$, satisfying $\tau_k^{\mathrm{sum}} = \tau_k^{\mathrm{s}} F_k^{\mathrm{s}} + \tau_k^{\mathrm{c}}$ and $\hat{f}_k^{\mathrm{e}} = f_k^{\mathrm{e}}$ for $k=1,\ldots,K$. Then, constraint~\eqref{eq:c4} can be rewritten as
\begin{equation}
T_k^{\mathrm{sen}} \!+\! T_k^{\mathrm{comp}} \!+\! T_k^{\mathrm{trans}} \!+\! {\tilde{N}_k T_k^{\mathrm{s}} F_k^{\mathrm{s}} C^{\mathrm{e}} ( 1 \!-\! p_{k,1}^{\mathrm{m}} )}/{\hat{f}_k^{\mathrm{e}}} \!\leq\! T_k^{\max}.
\label{eq:c4_new}
\end{equation}
Moreover, constraint~\eqref{eq:c6} becomes $\sum_{k=1}^{K} \tau_k^{\mathrm{sum}} \le 1$.

By dualizing the above two coupling equalities, problem~\eqref{eq:obj_fun} is decomposed into a device-side accuracy maximization subproblem and a global resource allocation subproblem. For device $k$, the subproblem is
\begin{subequations}
\begin{eqnarray}
\!\!&\!\!\!\!\!\!\!
\max\limits_{\left\{\!\substack{F_k^{\mathrm{s}}, \Delta n_k, \\
\delta_k, \tau_k^{\mathrm{c}}, \hat{f}_k^{\mathrm{e}}}\!\right\}}
&\!\!\!\!
p_{k,1}^{\mathrm{m}} \!\!+\!\! \sum_{i=2}^{I} p_{k,i}^{\mathrm{m}} \alpha_k (F_k^{\mathrm{s}}, \Delta n_k)
\!-\! \frac{\rho_2}{2} \Big\| f_k^{\mathrm{e}} \!\!-\!\! \hat{f}_k^{\mathrm{e}} \!\!+\!\! \frac{\beta_{2,k}}{\rho_2} \Big\|^2
\nonumber \\
&\!\!\!\!\!\!\!&\!\!\!\!
- \frac{\rho_1}{2} \Big\|\tau_k^{\mathrm{sum}} - (\tau_k^{\mathrm{s}} F_k^{\mathrm{s}} + \tau_k^{\mathrm{c}}) + \frac{\beta_{1,k}}{\rho_1} \Big\|^2,
\label{eq:subproblem_1}
~~~~~~\\
&\!\!\!\!\!\!\!\text{s.t.} &\!\!\!\!
\eqref{eq:c1}\text{--}\eqref{eq:c3},
\eqref{eq:c5},
\eqref{eq:c8}\text{--}\eqref{eq:c10},
\text{and}\,\eqref{eq:c4_new}.
\nonumber
\end{eqnarray}
\end{subequations}
The corresponding global resource allocation subproblem is
\begin{subequations}
\begin{eqnarray}
\!\!&\!\!\!\!\!\!\!
\min\limits_{\left\{\substack{\tau_1^{\mathrm{sum}},\ldots,\tau_K^{\mathrm{sum}}, \\
f_1^{\mathrm{e}},\ldots, f_K^{\mathrm{e}}}\right\}}
&\!\!\!\!
\frac{1}{K} \sum_{k=1}^{K} \frac{\rho_1}{2} \Big\| \tau_k^{\mathrm{sum}} - (\tau_k^{\mathrm{s}} F_k^{\mathrm{s}} + \tau_k^{\mathrm{c}}) + \frac{\beta_{1,k}}{\rho_1} \Big\|^2
\nonumber \\
&\!\!\!\!\!\!\!&\!\!\!\!
+ \frac{1}{K} \sum_{k=1}^{K} \frac{\rho_2}{2} \Big\| f_k^{\mathrm{e}} - \hat{f}_k^{\mathrm{e}} + \frac{\beta_{2,k}}{\rho_2} \Big\|^2 ,
\label{eq:subproblem_2}
~~~~~~\\
&\!\!\!\!\!\!\!\text{s.t.} &\!\!\!\!
\eqref{eq:c7},
\eqref{eq:c10},
\text{and } \sum_{k=1}^{K} \tau_k^{\mathrm{sum}} \le 1.
\nonumber
\end{eqnarray}
\end{subequations}
The dual variables are updated as
\begin{align}
\beta_{1,k}^{(i+1)}
&=
\beta_{1,k}^{(i)} \!+\! \rho_1 (\tau_k^{\mathrm{sum}} \!-\! (\tau_k^{\mathrm{s}} F_k^{\mathrm{s}} \!+\! \tau_k^{\mathrm{c}}) ), k = 1, \cdots, K,
\label{eq:beta_1_update}
\\
\beta_{2,k}^{(i+1)}
&=
\beta_{2,k}^{(i)} \!+\! \rho_2 (f_k^{\mathrm{e}} \!-\! \hat{f}_k^{\mathrm{e}}), k = 1, \cdots, K.
\label{eq:beta_2_update}
\end{align}

\subsection{Sensing Accuracy Maximization for Each Device}

We next simplify problem~\eqref{eq:subproblem_1} by transforming several coupled constraints. Since constraints~\eqref{eq:c1} and~\eqref{eq:c2} depend on both $F_k^{\mathrm{s}}$ and $\delta_k$, while $\delta_k$ only appears in these two constraints, they can be converted into an equivalent lower bound on $F_k^{\mathrm{s}}$. Specifically, for a given $F_k^{\mathrm{s}}$, the feasibility of constraints~\eqref{eq:c1} and~\eqref{eq:c2} can be checked by properly choosing $\delta_k$, which determines the tradeoff between the miss rate and the false positive rate. Following~\cite{he2024integrated}, increasing $F_k^{\mathrm{s}}$ improves this feasibility. Therefore, there exists a minimum feasible sampling rate, denoted by $F_k^{\mathrm{s,min}}$, which can be obtained offline by bisection search. As a result, constraints~\eqref{eq:c1} and~\eqref{eq:c2} are equivalently replaced by $F_k^{\mathrm{s}} \ge F_k^{\mathrm{s,min}}$.

Similarly, since $A_k(\Delta n_k)$ is monotonically decreasing with $\Delta n_k$, constraint~\eqref{eq:c3} is equivalent to $\Delta n_k \le \Delta n_k^{\max}$. Moreover, constraint~\eqref{eq:c5} can be equivalently transformed into $\Delta n_k \ge u(F_k^{\mathrm{s}})$, where $u(F_k^{\mathrm{s}})$ denotes the minimum subcarrier selection interval that satisfies the energy budget for a given $F_k^{\mathrm{s}}$. Its explicit form is omitted here due to the complexity of the expression. Then, problem~\eqref{eq:subproblem_1} can be reformulated as
\begin{subequations}
\label{pb:sp_1_new}
\begin{eqnarray}
\!\!&\!\!\!\!\!\!\!
\max\limits_{\left\{\!\substack{F_k^{\mathrm{s}}, \Delta n_k, \\
\tau_k^{\mathrm{c}}, \hat{f}_k^{\mathrm{e}} }\!\right\}} 
&\!\!\!\!\!\!\!
p_{k,1}^{\mathrm{m}} \!+\! \sum_{i=2}^{I} p_{k,i}^{\mathrm{m}} \alpha_k (F_k^{\mathrm{s}}, \Delta n_k ) \!- \! \frac{\rho_2}{2} \!\| f_k^{\mathrm{e}} \!\!-\!\! \hat{f}_k^{\mathrm{e}} \!\!+\!\! \frac{\beta_{2,k}}{\rho_2} \!\|^2 \nonumber \\
&\!\!\!\!\!\!\!&\!\!\!\!\!\!\!
- \frac{\rho_1}{2} \!\| \tau_k^{\mathrm{sum}}\!\! \!-\!\! (\tau_k^{\mathrm{s}} F_k^{\mathrm{s}} \!+\! \tau_k^{\mathrm{c}}) \!\!+\!\! \frac{\beta_{1,k}}{\rho_1} \|^2
\!\!\!,
\label{eq:subproblem_1_new_1}
~~~~~~\\
&\!\!\!\!\!\!\!\text{s.t.} &\!\!\!\!\!\!\!
\eqref{eq:c4_new}, 
F_k^{\mathrm{s}} \!\ge\! F_k^{\mathrm{s,min}}, F_k^{\mathrm{s}} \!\in\! \mathbb{Z}^{+}, \tau_k^{\mathrm{c}}, \hat{f}_k^{\mathrm{e}} \!\ge\! 0, \Delta n_k \!\in\! \mathbb{Z},
\nonumber \\
&\!\!\!\!\!\!\!&\!\!\!\!\!\!\!
u(F_k^{\mathrm{s}}) \!\le\! \Delta n_k \!\le\! \min(N\!-\!1,\Delta n_k^{\max}).
\end{eqnarray}
\end{subequations}

To solve~\eqref{pb:sp_1_new}, we relax $\Delta n_k$ to a continuous variable and simplify the ceiling operation in $\tilde N_k$ in~\eqref{eq:c4_new}. For a given $F_k^{\mathrm{s}}$, it is easy to prove that the resulting problem is convex with respect to $\Delta n_k$, $\tau_k^{\mathrm{c}}$, and $\hat{f}_k^{\mathrm{e}}$. However, since $\alpha_k(F_k^{\mathrm{s}},\Delta n_k)$ does not admit an explicit gradient, the Karush-Kuhn-Tucker (KKT) conditions cannot be directly applied. Therefore, we adopt a two-layer search procedure. In the outer layer, $F_k^{\mathrm{s}}$ is enumerated over $[F_k^{\mathrm{s,min}}, F_k^{\mathrm{s,max}}]$, where $F_k^{\mathrm{s,max}}$ is determined by the delay, energy, and time allocation constraints. In the inner layer, $\Delta n_k$ is optimized by golden-section search over $[u(F_k^{\mathrm{s}}), \Delta n_k^{\max}]$.

For each candidate pair $(F_k^{\mathrm{s}},\Delta n_k)$, the remaining variables $\tau_k^{\mathrm{c}}$ and $\hat{f}_k^{\mathrm{e}}$ are updated from the KKT conditions. If constraint~\eqref{eq:c4_new} is inactive, the optimal update is directly given by $\tau_k^{\mathrm{c},*} \!=\! \tau_k^{\mathrm{sum}} \!-\! \tau_k^{\mathrm{s}} F_k^{\mathrm{s}} \!+\! {\beta_{1,k}}/{\rho_1}$ and $\hat{f}_k^{\mathrm{e},*} \!=\! f_k^{\mathrm{e}} \!+\! {\beta_{2,k}}/{\rho_2}$. Otherwise, $\tau_k^{\mathrm{c},*}$ and $\hat{f}_k^{\mathrm{e},*}$ are determined by a one-dimensional bisection search over the associated dual variable until~\eqref{eq:c4_new} holds with equality. After the continuous solution is obtained, $\Delta n_k$ is rounded up to the nearest feasible integer satisfying~\eqref{eq:c4_new}, and the best objective value over all candidate $F_k^{\mathrm{s}}$ is selected.

\subsection{Edge Resource Allocation}

It is easy to prove that Problem~\eqref{eq:subproblem_2} is convex. Applying the KKT conditions yields
\begin{align}
&\tau_k^{\mathrm{sum},*}
=
(
(\tau_k^{\mathrm{s}} F_k^{\mathrm{s}} + \tau_k^{\mathrm{c}})
- \frac{\beta_{1,k}}{\rho_1}
- \frac{K\mu_1^*}{\rho_1}
)^+,
\label{eq:tau_k_sum_*}
\\
&f_k^{\mathrm{e},*}
=
(
\hat{f}_k^{\mathrm{e}}
- \frac{\beta_{2,k}}{\rho_2}
- \frac{K\mu_2^*}{\rho_2}
)^+,
\nonumber
\end{align}
where $(x)^+=\max(x,0)$, and $\mu_1^*$ and $\mu_2^*$ are chosen to satisfy $\sum_{k=1}^{K}\tau_k^{\mathrm{sum},*}\le 1$ and $\sum_{k=1}^{K}f_k^{\mathrm{e},*}\le f^{\mathrm{e}}$, respectively.

The update can be implemented as follows. We first set $\mu_1^*=\mu_2^*=0$. If the resulting $\tau_k^{\mathrm{sum},*}$ and $f_k^{\mathrm{e},*}$ already satisfy the two resource constraints, the update is completed. Otherwise, $\mu_1^*$ and $\mu_2^*$ are obtained by one-dimensional bisection search, with search intervals $0 \le \mu_1^* \le \frac{\rho_1}{K}\max_k\{(\tau_k^{\mathrm{s}} F_k^{\mathrm{s}} + \tau_k^{\mathrm{c}})-\beta_{1,k}/\rho_1\}$ and $0 \le \mu_2^* \le \frac{\rho_2}{K}\max_k\{\hat{f}_k^{\mathrm{e}}-\beta_{2,k}/\rho_2\}$. This yields the global time and computing resource allocation in each ADMM iteration.

\subsection{Overall Algorithm}

Algorithm~\ref{alg:admm} summarizes the overall ADMM-based procedure. In each iteration, each IoT device first solves~\eqref{eq:subproblem_1} using the above device-side search and update procedure. After that the edge server solves~\eqref{eq:subproblem_2} using the updates in~\eqref{eq:tau_k_sum_*} and the associated bisection search. The dual variables are then updated according to~\eqref{eq:beta_1_update} and~\eqref{eq:beta_2_update}. The iterations stop when the primal residuals fall below the threshold or the maximum iteration number is reached.

\begin{algorithm}[t]
\caption{ADMM-based Algorithm for Solving Problem~\eqref{eq:obj_fun}.}
\label{alg:admm}
\DontPrintSemicolon

\nlnonumber
\hspace*{-1em}\textbf{Initialize:} Set $\tau_k^{\mathrm{sum}}$, $f_k^{\mathrm{e}}$, $\beta_{1,k}$, $\beta_{2,k}$, $\rho_1$, $\rho_2$, primal residual tolerance $\varepsilon>0$, current iteration number $i=0$, and maximum iteration number $I_{\max}$;\;

\Repeat{$\|\tau_k^{\mathrm{sum},*}\!\!-\!(\tau_k^{\mathrm{s}} F_k^{\mathrm{s}} \!+ \!\tau_k^{\mathrm{c}})\|_2\le\varepsilon$, $\|f_k^{\mathrm{e},*}\!-\!\hat{f}_k^{\mathrm{e}}\|_2\!\le\!\varepsilon$, \textnormal{or} $i\!>\!I_{\max}$}{
    \% Accuracy maximization at the IoT devices;\;
    \For{$k=1,\ldots,K$}{
        Solve problem~\eqref{eq:subproblem_1} according to the above device-side procedure and obtain $F_k^{\mathrm{s}}$, $\Delta n_k$, $\tau_k^{\mathrm{c}}$, and $\hat{f}_k^{\mathrm{e}}$;\;
    }
    \% Resource allocation at the edge server;\;
    Solve problem~\eqref{eq:subproblem_2} according to the above edge-side update and obtain $\tau_k^{\mathrm{sum},*}$ and $f_k^{\mathrm{e},*}$;\;
    
    Update $\beta_{1,k}$ and $\beta_{2,k}$ according to \eqref{eq:beta_1_update} and \eqref{eq:beta_2_update};\;
    
    $i=i+1$;\;
}

Output the final solution to problem~\eqref{eq:obj_fun}.\;
\end{algorithm}

\section{Evaluation Results}
\label{sec:evalation}

In this section, we evaluate the proposed framework and algorithm through comprehensive experiments.

\begin{figure*}[t]
    \centering
    \includegraphics[width=0.98\linewidth]{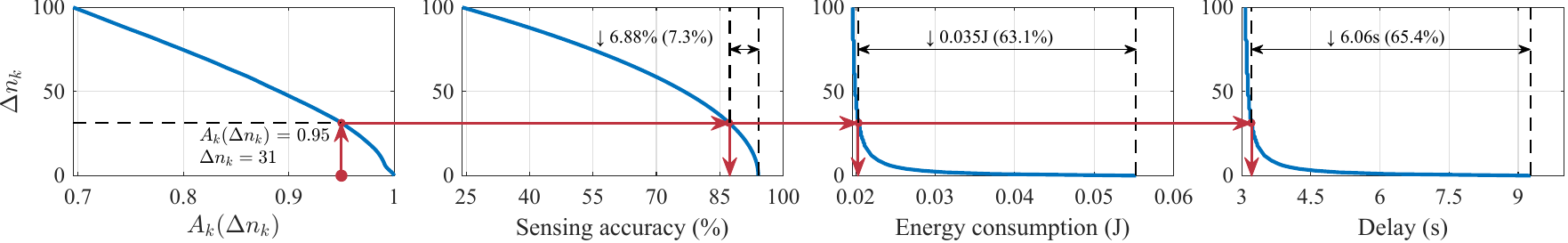}
    \vspace{-2.5ex}
    \caption{Impact of $\Delta n_k$ selected by the subcarrier similarity function on sensing accuracy, energy consumption, and delay.}
    \vspace{-2.5ex}
    \label{fig:single_device}
\end{figure*}

\begin{figure*}[t]
    \centering
    \begin{minipage}[t]{0.23\textwidth}
        \centering
        \includegraphics[width=\textwidth]{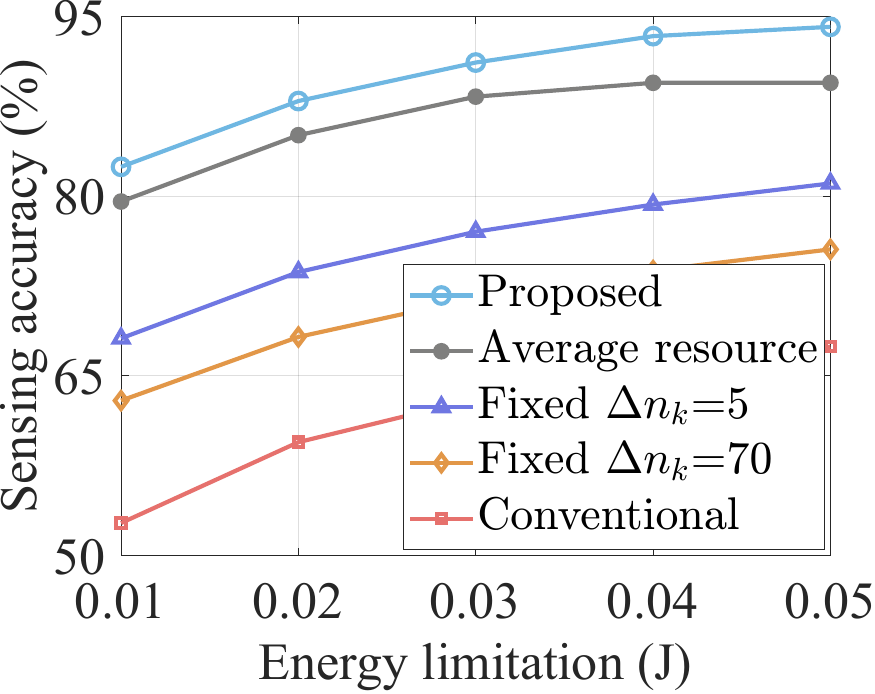}
        \vspace{-4.5ex}
        \caption{$E_k^{\max}$ vs. accuracy.}
        \label{fig:Performance_vs_E_max}
    \end{minipage}
    \hfill
    \begin{minipage}[t]{0.23\textwidth}
        \centering
        \includegraphics[width=\textwidth]{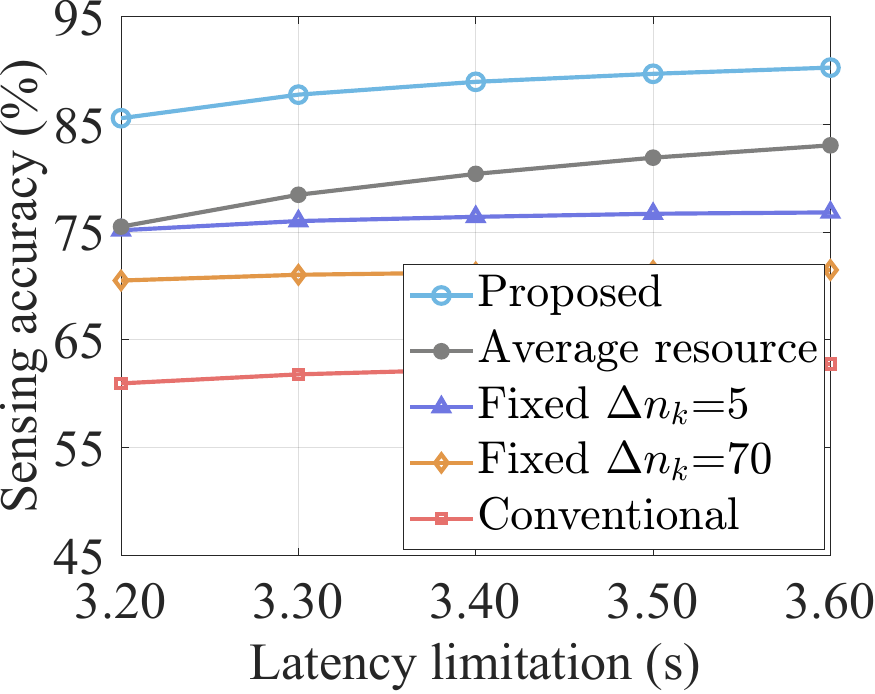}
        \vspace{-4.5ex}
        \caption{$T_k^{\max}$ vs. accuracy.}
        \label{fig:Performance_vs_T_max}
    \end{minipage}
    \hfill
    \begin{minipage}[t]{0.22\textwidth}
        \centering
        \includegraphics[width=\textwidth]{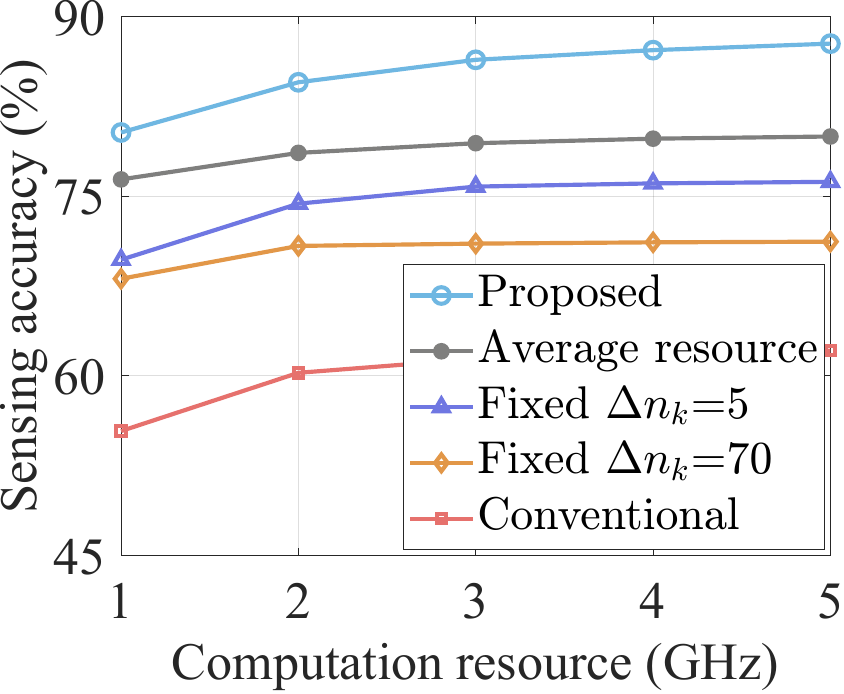}
        \vspace{-4.5ex}
        \caption{$f^{\mathrm{e}}$ vs. accuracy.}
        \label{fig:Performance_vs_f_e_total}
    \end{minipage}
    \hfill
    \begin{minipage}[t]{0.23\textwidth}
        \centering
        \includegraphics[width=\textwidth]{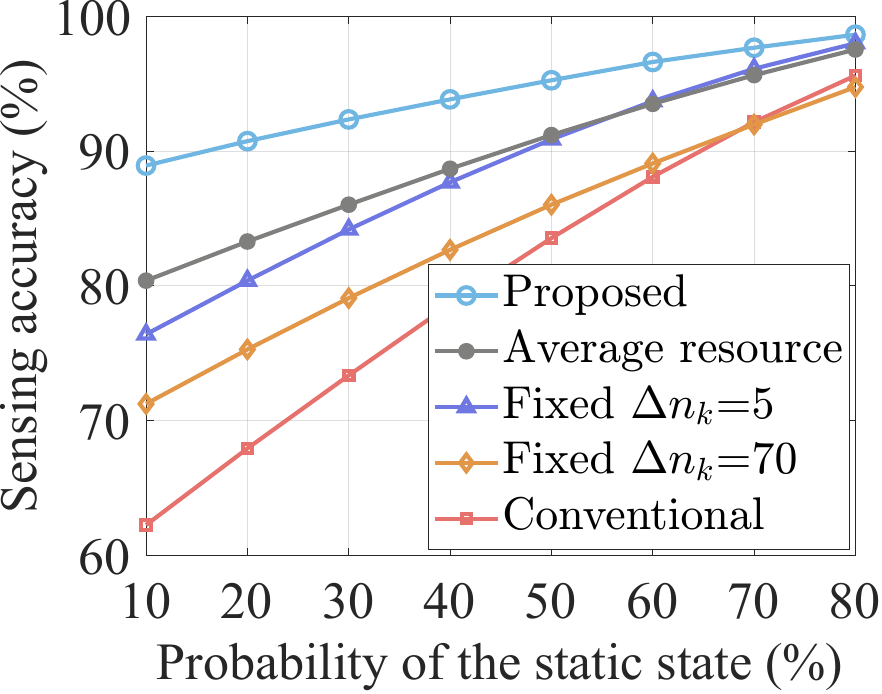}
        \vspace{-4.5ex}
        \caption{$p_{k,1}^{\mathrm{m}}$ vs. accuracy.}
        \label{fig:Performance_vs_p_m_1}
    \end{minipage}
    \vspace{-5ex}
\end{figure*}

\subsection{Evaluation Setup}
\label{sec:eva_set}
For evaluation, we use simulation for communication and real experiments for sensing. We consider a BS with an edge server and 12 randomly distributed IoT devices within a coverage radius of 550~\!m. The communication bandwidth is 20~\!MHz, with 245 subcarriers. The noise spectral density is -174~\!dBm/Hz. The communication transmit power, sensing power, and local computation power of each IoT device are set to 24.0~\!dBm, 20.4~\!dBm, and 25.4~\!dBm, respectively. Unless otherwise specified, $E_k^{\max}=0.03$~\!J, $T_k^{\max}=3.4$~\!s (including $T^{\mathrm{s}}=3$~\!s), $p^{\max}=2~\!\%$, $\eta_k=0.95$, and $f^{\mathrm{e}}=10$~\!GHz. For sensing, channel state information (CSI) data are collected in an office using commodity WiFi network interface cards (Intel AX210) under the IEEE 802.11ax standard. In this work, the measured CSI data are used as OFDM sensing data. The carrier frequency is 5~\!GHz, the bandwidth is 80~\!MHz, and the sampling rate is 2{,}000~\!Hz. Ten volunteers perform eight activity types, yielding 9{,}600 CSI time series. For each transmitter--receiver antenna pair, 245 adjacent subcarriers are selected from the measured CSI data to represent the sensing data of one IoT device under a 20~\!MHz bandwidth. This approximation is adopted because 245 adjacent subcarriers correspond to the subcarrier scale of a 20~\!MHz OFDM system. Based on these data, we fit $\lambda_i$, $r_i$, and $\sigma_\mathrm{c}^2$, and use a CNN to obtain $\alpha_k(F_k^{\mathrm{s}}, \Delta n_k)$. The probability of the static type is set to 0.1, and each of the other seven types has probability~0.9/7.

\subsection{Validation of the Subcarrier Selection Module}
Before evaluating the proposed framework and algorithm, we first validate the proposed subcarrier selection module on a single IoT device. Using the previously collected sensing dataset, we select the data of one IoT device and set $F_k^{\mathrm{s}}=500~\!$Hz. The results are shown in Fig.~\ref{fig:single_device}, with $\eta_k=0.95$. As $A_k(\Delta n_k)$ decreases from one to $\eta_k$, $\Delta n_k$ increases accordingly. The sensing accuracy decreases by only 7.3\%, while the energy consumption and delay are reduced by 63.1\% and 65.4\%, respectively. This indicates that the CSI time series across subcarriers contain substantial redundant information, and reducing the number of subcarriers can markedly lower energy consumption and delay with only a minor loss in sensing accuracy. Therefore, the proposed subcarrier selection module is effective in the single-device setting.

\subsection{Performance Comparison}
To further evaluate the proposed framework and algorithm, we consider three baselines: 1) \emph{Conventional scheme}~\cite{he2024integrated}: no subcarrier selection is applied. 2) \emph{Fixed $\Delta n_k$ scheme}: $\Delta n_k$ is fixed, with $\Delta n_k \in \{5,70\}$. 3) \emph{Average resource scheme}: time and edge computing resources are equally allocated among IoT devices, i.e., $(\tau_k^{\mathrm{sum}}, f_k^{\mathrm{e}}) = (1/K,\, f^{\mathrm{e}}/K)$.

First, we evaluate the impact of different $E_k^{\max}$ on the four schemes, as shown in Fig.~\ref{fig:Performance_vs_E_max}. As $E_k^{\max}$ increases, the sensing accuracy of all schemes improves, since a larger energy budget allows a higher $F_k^{\mathrm{s}}$ or a smaller $\Delta n_k$, both of which improve $\alpha_k(F_k^{\mathrm{s}}, \Delta n_k)$. The proposed scheme consistently outperforms the other schemes under different energy budgets, because it can adaptively optimize $\Delta n_k$ for different IoT devices and allocate resources more efficiently. By contrast, the conventional scheme performs the worst because it offloads complete CSI data and thus supports only a relatively low feasible sampling rate under limited resources. The fixed $\Delta n_k$ scheme performs better than the conventional one, but remains inferior to the proposed scheme due to its lack of flexibility. In particular, the case $\Delta n_k=5$ generally outperforms $\Delta n_k=70$, indicating that an overly large fixed interval may lead to excessive sensing information loss.

Fig.~\ref{fig:Performance_vs_T_max} illustrates the impact of different $T_k^{\max}$ on the four schemes. Similar to the case with limited energy, the sensing accuracy of all schemes improves as $T_k^{\max}$ increases, because a looser delay constraint allows a higher $F_k^{\mathrm{s}}$ or a smaller $\Delta n_k$. The proposed scheme again consistently achieves the best performance, showing its advantage under different delay budgets. Meanwhile, the performance gains of the conventional scheme and the fixed $\Delta n_k$ scheme are less significant than those of the proposed scheme and the average resource scheme. This suggests that, under the current evaluation settings, delay is not the primary bottleneck for the conventional and fixed $\Delta n_k$ schemes.

Fig.~\ref{fig:Performance_vs_f_e_total} shows the impact of the total computation resource $f^{\mathrm{e}}$ on the four schemes. As $f^{\mathrm{e}}$ increases, the sensing accuracy of all schemes improves because more edge computing resources reduce the recognition delay and enable more favorable sensing configurations. The proposed scheme consistently outperforms the baselines, further confirming its effectiveness. It is also observed that the performance of all schemes first improves significantly and then gradually saturates as $f^{\mathrm{e}}$ increases. This is because, once $f^{\mathrm{e}}$ becomes sufficiently large, edge computing is no longer the dominant factor limiting the sensing accuracy.

Fig.~\ref{fig:Performance_vs_p_m_1} illustrates the impact of the probability of the static type $p_{k,1}^{\mathrm{m}}$ on the four schemes. As $p_{k,1}^{\mathrm{m}}$ increases, the sensing accuracy of all schemes improves, and the proposed scheme consistently achieves the best performance. However, the performance gap among the schemes gradually narrows as $p_{k,1}^{\mathrm{m}}$ increases. This is because the miss rate and false positive rate are constrained to be close to zero, so the static type can be recognized correctly. As a result, when static samples become dominant, the accuracies of all schemes increase and become closer to one another.

\section{Conclusion}
\label{sec:con}

In this paper, we propose a subcarrier selection-based sensing framework for multi-device ISCC systems to reduce sensing overhead by removing frequency-domain redundancy from OFDM sensing data. We establish analytical models for sensing accuracy, delay, and energy consumption, formulate a sensing accuracy maximization problem under practical constraints, and develop an ADMM-based algorithm for joint sensing parameter optimization and resource allocation. Experiment results on commodity wireless devices validate the proposed framework and show that it consistently outperforms three baseline schemes under different resource constraints. This work demonstrates the potential of subcarrier selection for improving sensing efficiency in future ISCC systems.

\bibliographystyle{IEEEtran}
\bibliography{ref}

\end{document}